\documentclass[fleqn,usenatbib]{mnras}

\usepackage{newtxtext,newtxmath}

\usepackage[T1]{fontenc}
\usepackage{ae,aecompl}

\usepackage{amsmath, amssymb}
\usepackage{amsfonts}
\usepackage{epsfig,rotating}

\def\simeq{
\mathrel{\raise.3ex\hbox{$\sim$}\mkern-14mu\lower0.4ex\hbox{$-$}}
}

\def\ltsima{$\; \buildrel < \over \sim \;$}
\def\simlt{\lower.5ex\hbox{\ltsima}}
\def\gtsima{$\; \buildrel > \over \sim \;$}
\def\simgt{\lower.5ex\hbox{\gtsima}}

\def\msun{{\rm M_{\odot}}}

\def\be{\begin{equation}}
\def\ee{\end{equation}}

\def\del#1{{}}
\def\ltsima{$\; \buildrel < \over \sim \;$}
\def\simlt{\lower.5ex\hbox{\ltsima}}
\def\gtsima{$\; \buildrel > \over \sim \;$}
\def\simgt{\lower.5ex\hbox{\gtsima}}

\title[Warm absorbers: supermassive black hole feeding, and Compton--thick AGN]{Warm absorbers: supermassive black hole feeding,
and Compton--thick AGN}

\author[Kastytis Zubovas and Andrew King]{Kastytis Zubovas$^{1,2,\star}$ and Andrew King$^{3,4,5}$ \\
  $^{1}$Center for Physical Sciences and Technology, Savanori\c{u} 231, Vilnius LT-02300, Lithuania \\
  $^{2}$Vilnius University Observatory, Saulėtekio av. 3, Vilnius LT-10257, Lithuania \\
  $^{3}$Department of Physics \& Astronomy, University of Leicester, Leicester, LE1 7RH, UK \\
  $^{4}$ Astronomical Institute Anton Pannekoek, University of Amsterdam, Science Park 904, 1098 XH Amsterdam, Netherlands\\ 
  $^{5}$ Leiden Observatory, Leiden University, Niels Bohrweg 2, NL-2333 CA Leiden, Netherlands \\
  $^{\star}$ {E-mail:~} {\rm kastytis.zubovas@ftmc.lt} }

\date{Accepted XXX. Received YYY; in original form ZZZ}

\pubyear{2018}

\begin{document}
\label{firstpage}
\pagerange{\pageref{firstpage}--\pageref{lastpage}}
\maketitle

\begin{abstract}
Warm absorbers are found in many AGN and consist of clouds moving at moderate radial velocities, showing complex ionization structures and having moderate to large column densities. Using 1D numerical calculations, we confirm earlier suggestions that the energy released by an AGN pushes the surrounding gas outward in a bubble until this reaches transparency. Typical AGN episode durations of $5\times 10^4$~yr supply enough energy for this, except in very gas-rich and/or very compact galaxies, such as those in the early Universe. In those galaxies, the AGN might remain hidden for many periods of activity, hiding the black hole growth. The typical radii of $0.1-1$~kpc, velocities of $100-1000$~km~s$^{-1}$ and resulting optical depths are consistent with observations of warm absorbers. The resulting structure is a natural outcome of outflows driven by AGN buried in an optically thick gas envelope, and has a total mass comparable to the final $M -\sigma$ mass the central supermassive black hole will eventually reach.These results suggest that AGN can feed very efficiently by agitating this surrounding dense material. This may not be easy to observe, as this gas is Compton thick along many sightlines. The infall may produce episodic star formation in the centre, building up nuclear star clusters simultaneously with the growth of the central black hole.
\end{abstract}



\section{Introduction}

Supermassive black holes (SMBHs), with masses $M_{\rm BH}$ ranging from a few times $10^4 \; \msun$ to more than $10^{10} \; \msun$, are present in the vast majority of galaxies and can affect their evolution in numerous ways \citep[e.g.,][]{Cattaneo2009Natur,King2015ARA&A,Harrison2018NatAs}. During episodes of rapid accretion, active galactic nuclei (AGN) reach luminosities $L \sim L_{\rm Edd}$, with $L_{\rm Edd} = 4 \pi G M_{\rm BH} c / \kappa$ the Eddington luminosity. This luminosity can easily exceed the total stellar luminosity of the host galaxy and drive massive outflows that remove significant amounts of gas, possibly quenching star formation \citep{Feruglio2010A&A,Sturm2011ApJ,Cicone2015A&A}. Outflows are detected in a majority, $\sim 60\%$, of AGN \citep{Ganguly2008ApJ}.

There are several theoretical models attempting to explain  outflow driving, such as AGN jets \citep[e.g.][]{Wagner2012ApJ} and radiation pressure \citep{Ishibashi2014MNRAS}. One of the most promising models is that of radiation-driven relativistic AGN winds \citep{King2003MNRASb,King2003ApJ}, which are observed in $\sim 40\%$ of AGN \citep{Tombesi2010A&A,Tombesi2010ApJ}. These winds can drive massive outflows \citep{King2010MNRASa} and clear gas out of galaxies \citep{Zubovas2012ApJ}, compress gas and trigger starbursts \citep{Zubovas2016MNRASb} and establish the $M - \sigma$ relation \citep{King2003ApJ}. Nevertheless, distinguishing among these models and their predictions is not generally easy, and more observationally testable predictions are required.

One feature of AGN which may provide such testing grounds is the presence of warm absorbers (WAs). These are clouds of moderately-ionized gas moving with velocities of several hundred to a few thousand km/s, visible as absorption features in more than 50\% of AGN spectra \citep{Reynolds1997MNRAS,Crenshaw2003ARA&A,Porquet2004A&A,Laha2014MNRAS}. They show moderate to large column densities ($20 \simlt {\rm log} \; N_{\rm H} \simlt 23$~cm$^{-2}$) and ionization parameters ($-1 \simlt {\rm log} \; \xi \simlt 3$) \citep{Reeves2013ApJ}, with a positive correlation between ionization and velocity \citep{Pounds2013MNRAS}. Their existence may be a result of AGN radiation pressure pushing the surrounding material outward to the transparency radius. At this radius, the driving force decreases rapidly, and the shell stalls, providing a barrier for future winds to shock against \citep{King2014MNRAS}. These shocks continuously reheat the material, which then cools rapidly and produces the numerous ionization species observed in WAs. The total mass of the WA structure is comparable with the $M - \sigma$ mass the black hole will ultimately reach \citep{King2014MNRAS}.  

In this paper, we numerically analyse the dynamics of outflows driven by the total AGN radiation pressure absorbed in the optically thick gas shell. We investigate the importance of the initial gas mass of the shell and determine the salient properties of the outflows. We show that outflows of this type naturally tend to stall at around the transparency radius, even if the AGN shuts down long before the outflow reaches this extent. The properties of stalled outflows are close to observed warm absorber properties and depend only weakly on initial conditions and gas density profile. They depend, however, on total gas mass, suggesting that correlations between gas content in the galaxy and WA radius should exist. In addition, by clearing the surrounding gas reservoir, the AGN limits its activity episode duration and maximum luminosity before it becomes visible to outside observers; therefore, multiple AGN episodes in a single galaxy should all have similar properties.

The paper is structured as follows. In Section \ref{sec:wind} we describe the outflow model in some detail, in particular considering the obscuration of the AGN in the initial phases of outflow expansion. We present the setup of our numerical calculations in Section \ref{sec:numerics}. In Section \ref{sec:was} we present the main results of outflow expansion and WA formation. Finally, we discuss the implications of our results in Section \ref{sec:discuss} and conclude in Section \ref{sec:concl}.

\section{The outflow model} \label{sec:wind}

Our model is based very closely on the AGN wind-driven feedback model, which has been very successful in explaining the properties of large-scale AGN-driven outflows \citep{Zubovas2012ApJ}. The origin of the outflow is a quasi-relativistic wind launched from the accretion disc \citep{King2003ApJ}. The wind self-regulates to keep its Compton scattering optical depth $\tau \sim 1$ \citep{King2010MNRASb}, and therefore its momentum rate is 
\begin{equation} \label{eq:mom}
\dot{M}_{\rm w} v_{\rm w} = \frac{L_{\rm AGN}}{c},
\end{equation}
where $\dot{M}_{\rm w}$ is the wind mass flow rate, $v_{\rm w}$ is the wind velocity and $L_{\rm AGN}$ is the AGN luminosity; $c$ is the speed of light. From this, we derive the wind velocity
\begin{equation}
v_{\rm w} = \frac{\eta  c}{\dot{m}},
\end{equation}
where $\eta \simeq 0.1$ is the radiative efficiency of accretion and $\dot{m} \equiv \dot{M}_{\rm w}/\dot{M}_{\rm accr}$ is the ratio between wind mass flow rate and SMBH accretion rate. Observations indicate that wind velocities range from $\sim 0.03 c$ to $\sim 0.3 c$ \citep{Tombesi2012MNRAS,Tombesi2014MNRAS}, indicating a range of value for $\dot{m}$ \citep[see also][]{Tombesi2013MNRAS}. Importantly, the single-scattering relation (eq. \ref{eq:mom}) is valid for essentially all observed UFOs \citep[e.g.][]{Pounds2003MNRASa, Tombesi2015Natur, Chartas2018arXiv}. Winds with high values of $\dot{m}$ are slower and less energetic; in cases of extremely high $\dot{m}$, they may switch to broad absorption line (BAL) outflows \citep{Zubovas2013ApJ}. Other than this situation, the value of $\dot{m}$ does not have qualitative effects on wind and/or outflow behaviour. Therefore, for the sake of simplicity, we take $\dot{m} = 1$ in subsequent calculations. The kinetic energy rate of such a wind is
\begin{equation}
\dot{E}_{\rm w} = \frac{\dot{M}_{\rm w} v_{\rm w}^2}{2} = \frac{\eta}{2} L_{\rm AGN} \simeq 0.05 L_{\rm AGN}.
\end{equation}

When the wind encounters the surrounding interstellar medium (ISM), a strong shock front develops and an outflow bubble is formed. The subsequent evolution of the system depends on two factors: cooling of the shocked wind and opacity of the outflow shell. If the shell is transparent ($\tau_{\rm e.s.} < 1$) and the shocked wind cools efficiently, a momentum-driven outflow develops, with energy rate $\dot{E}_{\rm out} \ll 0.05 L_{\rm AGN}$, which establishes the $M-\sigma$ relation \citep{King2003ApJ}. If the shell is transparent but the shocked wind does not cool, most of the wind energy is transferred to the ISM and an energy-driven outflow develops, with $\dot{E}_{\rm out} \simeq 0.05 L_{\rm AGN}$, which clears gas out of the galaxy and quenches subsequent star formation \citep{Zubovas2012ApJ}. 

Two other outflow possibilities exist, both of which can lead to the outflow having a much higher total energy, $\dot{E}_{\rm out} \simeq L_{\rm AGN}$. In one of them, the wind is driven by radiation pressure, as above, but the outflowing shell is opaque to the AGN radiation. In this way, most of the AGN radiation becomes trapped inside the outflow and pushes it outward. This is not a particularly realistic scenario, since real outflows are clumpy rather than spherically symmetric, therefore there will almost always be some photon leakage. In addition, the AGN photons absorbed by the gas may be reradiated at lower energies, where opacity is lower. Nevertheless, the outflow can be effectively driven by a luminosity significantly higher than $0.05 L_{\rm AGN}$. The second possibility occurs if the black hole is embedded in a very dense gas reservoir or is fed by dense massive gas streams. In this case, the accretion rate can be super-Eddington, and all AGN photons are well mixed with the gas. The resulting wind velocity is not significantly increased, $v_{\rm w} \simlt 0.2c$, decreasing with increasing accretion rate \citep{King2016MNRAS}. This situation seems applicable to ultra-luminous X-ray sources \citep{Middleton2014MNRAS,Middleton2015MNRAS,Middleton2016MNRAS}, but is probably uncommon in AGN, where the accretion rate is hardly ever super-Eddington \citep{Jones2016ApJ}. It may be relevant at early cosmic times, however, when the SMBHs have not yet grown to their limiting masses as predicted by the $M-\sigma$ relation. We return to this point in Section \ref{sec:feeding}. In this case, again assuming that there is no avenue for significant photon leakage, the wind itself has a kinetic energy rate $\dot{E}_{\rm w} \simeq L_{\rm AGN}$. As this wind shocks against the surrounding ISM, provided that cooling is inefficient, we get $\dot{E}_{\rm out} \simeq \dot{E}_{\rm w} \simeq L_{\rm AGN}$, i.e. the same situation as in the case of a momentum-conserving wind hitting an optically thick gas envelope. In this paper, we are interested in the luminosity that drives the outflows; therefore, the cases of momentum-conserving wind and a wind driven by hyper-Eddington inflow are equivalent for our purposes. In such an outflow, radiation pressure dominates over gas pressure, leading to a change in the effective adiabatic index: $\gamma = 4/3$.

As mentioned above, gas opacity, and hence the optical depth of the shell, depends on wavelength; it may be that the shell is opaque to UV radiation but transparent to IR. If the UV photons are absorbed by dust, which then radiate in the IR, the coupling between the AGN radiation field and the gas envelope is reduced to approximately the momentum-driven case $\dot{M}_{\rm out}v_{\rm out} \simeq {L_{\rm AGN}}/c$ \citep{Ishibashi2012MNRAS,Ishibashi2018MNRAS}.

If we assume that the outflowing shell is spherically symmetric, its optical depth is given by
\begin{equation}
\tau_{\rm out} = \frac{\kappa M_{\rm out}\left(<R\right)}{4\pi R^2},
\end{equation}
where $M_{\rm out}\left(<R\right)$ is the total gas mass contained within radius $R$ which the outflow has reached. The radius of transparency, defined by the condition $\tau_{\rm out}\left(R_{\rm tr}\right) = 1$, is, for an isothermal gas distribution,
\begin{equation}
R_{\rm tr} = \frac{\kappa f_{\rm g} \sigma^2}{2\pi G},
\end{equation}
with $f_{\rm g}$ the gas fraction in the galaxy, $\sigma$ the velocity dispersion in the host galaxy spheroid and $G$ the gravitational constant. The mass of the outflowing material collected at the moment that the outflow becomes transparent has a simple relationship with the critical SMBH mass required for the removal of gas to large distances by AGN wind momentum alone, $M_\sigma = f_{\rm c} \kappa \sigma^4 / \pi G^2$, with $f_{\rm c} = 0.16$ the cosmological gas fraction \citep[cf.][]{King2010MNRASa,King2014MNRAS}:
\begin{equation} \label{eq:ratio_transp}
\frac{M_{\rm out}\left(\tau_{\rm out}=1\right)}{M_\sigma} = \frac{\pi G^2 M_{\rm out}}{f_{\rm c} \kappa \sigma^4} = \frac{\pi G^2 M_{\rm out}}{f_{\rm c} \kappa}\frac{4R^2}{G^2 M_{\rm tot}^2} = \frac{4 \pi R^2 f_{\rm g}}{\kappa f_{\rm c} M_{\rm tot}}=\frac{f_{\rm g}^2}{f_{\rm c}}.
\end{equation}
Here, we used the usual definition of velocity dispersion $\sigma^2 \equiv G M_{\rm tot} / 2 R$, with $M_{\rm tot} \equiv M_{\rm out} / f_{\rm g}$ the total mass contained within $R$. This expression for $\sigma$ is valid for any density profile, with the caveat that $\sigma = \sigma\left(R\right)$ except in an isothermal profile. If $f_{\rm g} = f_{\rm c}$, i.e. the galaxy has not been cleared yet, the mass of the outflowing shell at the moment of transparency should be equal to $f_{\rm g} M_\sigma$. In more gas-rich galaxies (e.g. after a major merger), the outflow mass is higher, and in gas-poor galaxies it is lower. The more massive the enshrouding envelope, the more energy is required to reach transparency; black holes embedded in dense gas envelopes might be unable to clear the surrounding gas even by these very energetic outflows. Assuming that the outflowing shell at the moment of transparency is observed as a warm absorber, we predict there should be a direct, super-linear correlation between WA mass and the total gas content in the galaxy: $M_{\rm WA} \propto M_{\rm g}^2$, where $M_{\rm g} = f_{\rm g} M_{\rm gal}$ is the gas mass in the whole galaxy of mass $M_{\rm gal}$. This correlation can be understood as a simple consequence of two relations: $M_{\rm WA} \simeq M_{\rm out} = f_{\rm g} M_{\rm tot}\left(<R\right)$ and $R \propto f_{\rm g}$. Both relations are not exact except for an isothermal profile, therefore we expect significant scatter around this prediction.

After an extensive literature search, we have been unable to find any prior work looking for correlations between properties of WAs and properties of their host galaxies. Such research would be a powerful test of our model.

In order to determine other salient properties of WAs within our model, we now turn to numerical calculations.

\section{Numerical method} \label{sec:numerics}
\subsection{Equation of motion}

We derive the equation of motion for a spherically symmetric gas shell with arbitrary adiabatic index $\gamma$, expanding in a spherically symmetric background potential, driven by a central power source injecting a kinetic power $L$. The derivation is based on that presented in \citet{King2005ApJ} and developed in \citet{King2014MNRAS} and \citet{Zubovas2016MNRASb}, so we refer the reader to those papers for details. We start by writing an appropriate expression of Newton's second law:
\begin{equation}
\frac{{\rm d}}{{\rm d}t}\left[M\dot{R}\right] = 4\pi R^2 P - \frac{GM\left[M_{\rm p}+M/2\right]}{R^2}.
\end{equation}
Here the term on the left hand side is the time derivative of the linear momentum, with $M\left(R\right)$ the instantaneous swept--up gas mass being driven out when the bubble (contact discontinuity between the shocked wind and shocked ISM) is at radius $R$, $P$ is the expanding gas pressure and $M_{\rm p}$ is the mass of the stars and dark matter within $R$. To this we add the energy equation
\begin{equation}
\frac{{\rm d}}{{\rm d}t}\left[\frac{3}{2}PV\right] = \frac{\eta}{2}L - P\frac{{\rm d}V}{{\rm d}t} - \frac{{\rm d}E_{\rm g}}{{\rm d}t},
\end{equation}
where the term on the left hand side is the change in wind internal energy, $V$ is the volume cleared by the outflowing gas and $E_{\rm g}$ is the gravitational binding energy of the gas. By writing the pressure and volume of the outflowing gas in terms of $R$, expanding all the derivatives and rearranging, we are finally left with an equation
\begin{equation} \label{eq:eom}
\dddot{R} = \frac{3 \left(\gamma-1\right)}{M_{\rm g} R} \left(\eta L - A\right) - B,
\end{equation}
where
\begin{equation}
\begin{split}
A &= \dot{M}_{\rm g} \dot{R}^2 + M_{\rm g} \dot{R} \ddot{R} + \frac{2 G\dot{R}}{R^2}\left(M_{\rm g}M_{\rm p} + \frac{M_{\rm g}^2}{2}\right) - \\ &-G\frac{\dot{M}_{\rm g}M_{\rm p}+M_{\rm g}\dot{M}_{\rm p}+M_{\rm g}\dot{M}_{\rm g}}{R}
\end{split}
\end{equation}
and
\begin{equation}
\begin{split}
B &= \frac{\ddot{M}_{\rm g}\dot{R}}{M_{\rm g}} + \frac{\dot{M}_{\rm g}\dot{R}^2}{M_{\rm g}R} + \frac{2 \dot{M}_{\rm g}\ddot{R}}{M_{\rm g}} + \frac{\dot{R}\ddot{R}}{R} + \\&+ G\frac{\dot{M}_{\rm g}M_{\rm p}+M_{\rm g}\dot{M}_{\rm p}+M_{\rm g}\dot{M}_{\rm g}}{M_{\rm g}R^2} - G\frac{2M_{\rm g}M_{\rm p}\dot{R}+M_{\rm g}^2 \dot{R}}{2M_{\rm g}R^3}.
\end{split}
\end{equation}
In these equations, $\dot{M}\equiv \dot{R}\partial M/\partial R$ and $\ddot{M} \equiv \ddot{R}\partial M/\partial R + \dot{R} ({\rm d}/{\rm d}t)\left(\partial M/\partial R\right)$. As long as $M_{\rm g}\left(R\right)$ and $M_{\rm p}\left(R\right)$ and their first and second derivatives with respect to $R$ are analytical, the value of $\ddot{R}$ can be calculated and the motion of the outflow integrated numerically.

\subsection{Simulation setup}

We integrate the equation of motion (eq. \ref{eq:eom}) numerically, using a one-dimensional KDK leapfrog scheme \citep[there is no appreciable difference between using this or a different integration scheme, cf.][]{Zubovas2016MNRASb}. We use initial conditions consisting of a galaxy with total mass $10^{12} \; \msun$, consisting of an NFW halo with mass $M_{\rm h} = 6\times 10^{11} \; \msun$ and gas fraction $f_{\rm g,h} = 10^{-3}$ and an isothermal bulge with mass $M_{\rm b} = 4 \times 10^{10} \; \msun$ and radius $R_{\rm b} = 1$~kpc. These bulge properties give a bulge velocity dispersion $\sigma_{\rm b} \simeq 293$~km~s$^{-1}$, which is rather large, but probably appropriate for high-redshift galaxies \citep{vanDokkum2009Natur, Trujilo2010IAUS}. We consider five values for the bulge gas fraction: $f_{\rm g,b} = 0.05, 0.1, 0.25, 0.5, 1$. We tested the calculations with Jaffe and NFW gas density profiles and found that the results remain essentially unchanged.

The mass of the central black hole is $M_{\rm BH} = 2 \times 10^8 \; \msun$. The AGN luminosity during the activity episode is $L_{\rm AGN} = L_{\rm Edd} = 2.6 \times 10^{46}$~erg~s$^{-1}$. The driving luminosity of the outflow is related to the AGN luminosity and the outflowing shell optical depth in the following way:
\begin{equation}
L = \begin{cases}
    \; L_{\rm AGN}, & \tau \geq 1, \\
    \; \tau L_{\rm AGN}, & 0.05 < \tau < 1, \\
    \; 0.05 L_{\rm AGN}, & \tau \leq 0.05.
  \end{cases}
\end{equation}

We run the calculations for a case of continuous AGN activity, and a case where the AGN switches off after $t_{\rm q} = 5\times 10^4$~yr, to account for the expected short lifetimes of AGN episodes \citep{King2015MNRAS, Schawinski2015MNRAS}. We are primarily interested in the conditions of the outflow at the moment when the AGN becomes visible, i.e. the optical depth of the outflow drops to $\tau < 1$. In addition, we investigate the coasting of outflows after the AGN switches off, to understand the structure of the SMBH surroundings by the time a new episode might occur.

\section{Evolution of optically thick outflows} \label{sec:was}
\subsection{Continuous outflow propagation}

\begin{figure}
	\includegraphics[width=0.97\columnwidth]{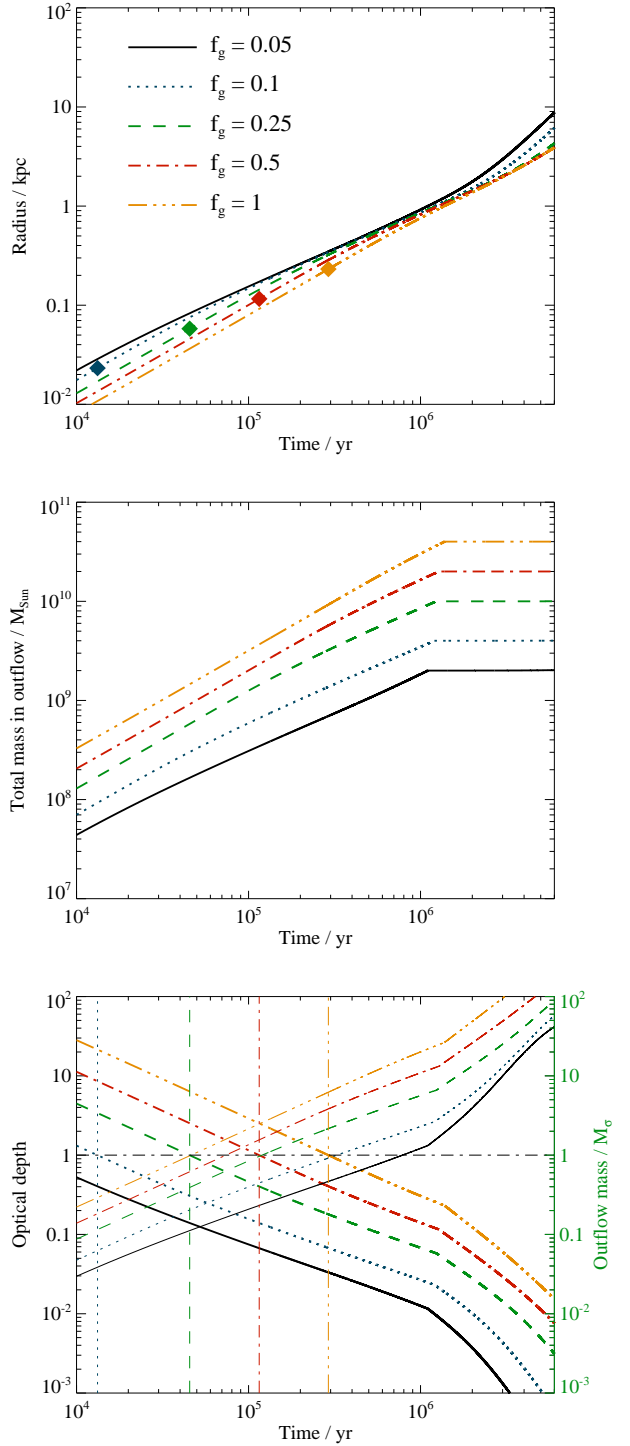}
    \caption{Evolution of outflow radius (top), total outflowing mass (middle) and optical depth and outflow mass ratio with $M_\sigma$ (bottom, thick and thin lines, respectively) for isothermal gas distributions with different total gas mass. The diamond symbols in the top panel show the location of the transparency radius. In the bottom panel, the horizontal line marks the transparency condition, and vertical lines show the times when simulations reach transparency. Sudden changes in total mass and mass ratio occur when the outflow escapes the isothermal bulge and starts expanding rather more freely through the halo, leading to a decrease in effective $\sigma$.}
    \label{fig:wabs_1}
\end{figure}

Figure \ref{fig:wabs_1} shows the salient outflow properties: radius against time (top), total mass against time (middle) and optical depth (bottom, thick lines); the ratio $M_{\rm out}/M_\sigma$ is shown in the bottom panel in thin lines. Additionally, we mark the transparency condition $\tau = 1$ with a horizontal line and the times when the models reach transparency with vertical lines. The five cases with different gas densities are shown with different line colours and styles, as indicated in the top panel. The outflow evolves in a qualitatively similar fashion for all values of gas density. Transparency is achieved in $t < 10^4$~yr in the least dense case and at $t \simeq 3\times 10^5$~yr in the densest one. The radius of transparency is similarly dependent on gas density, ranging from $R_{\rm tr} \sim 10$~pc to $R_{\rm tr} \sim 200$~pc from the least to most dense cases. The total outflowing mass varies much more, with $M_{\rm tr} \sim 3 \times 10^7 \; \msun$ in the least dense case, while the densest case has $M_{\rm tr} \sim 10^{10} \; \msun$. The ratio $M_{\rm out}/M_\sigma$ at the moment of transparency is equivalent to the analytical prediction (eq. \ref{eq:ratio_transp}).

\subsection{Coasting solutions} \label{sec:coasting}

\begin{figure}
	\includegraphics[width=\columnwidth,trim={0 0 0 0},clip]{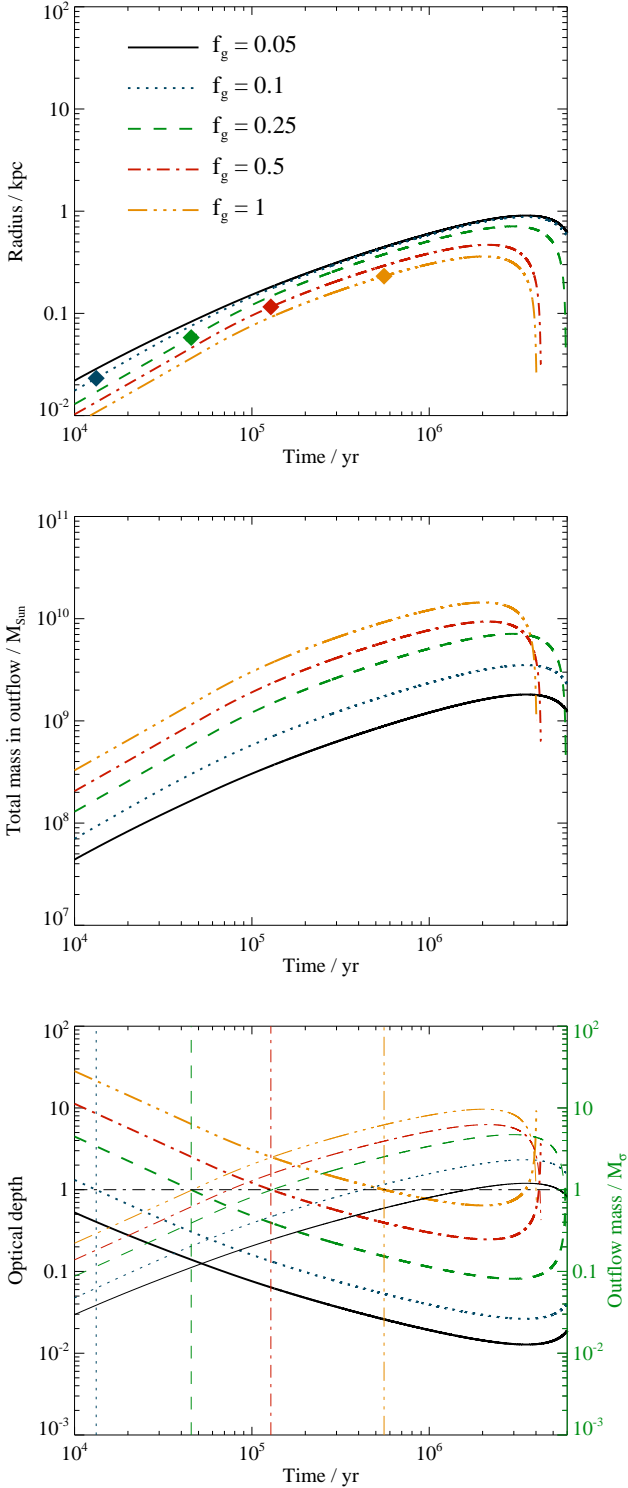}
    \caption{Same as Figure \ref{fig:wabs_1}, but with AGN episode duration limited to $5\times10^4$~yr. The lines are terminated at the position of the first ``adiabatic bounce'', which is an unrealistic consequence of our assumption of a perfectly adiabatic system.}
    \label{fig:coasting}
\end{figure}

\begin{figure}
	\includegraphics[width=\columnwidth,trim={0 0 0 0},clip]{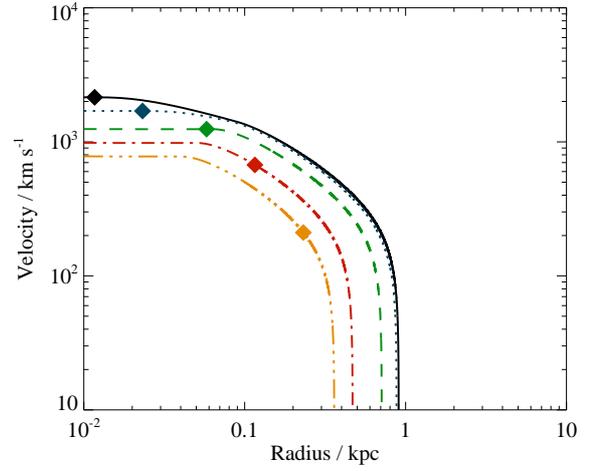}
    \caption{Outflow velocity as function of radius in simulations with AGN activity limited to a period of $5\times10^4$~yr. Diamond symbols show outflow velocity at the transparency radius.}
    \label{fig:vel_radius}
\end{figure}

\begin{figure}
	\includegraphics[width=\columnwidth,trim={0 0 0 0},clip]{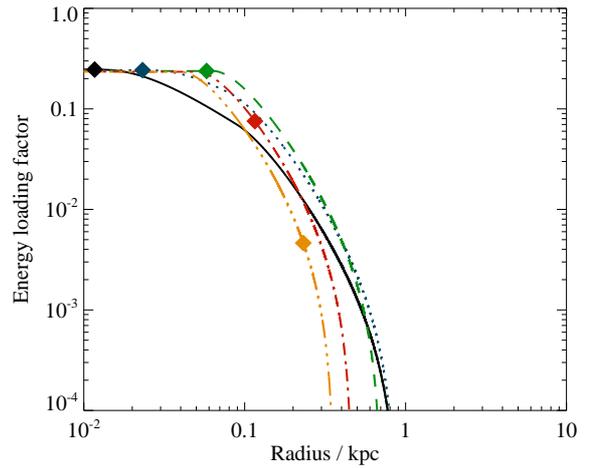}
    \caption{Effective energy loading factor as function of radius in simulations with AGN activity limited to a period of $5\times10^4$~yr. Diamond symbols show energy loading factor at the transparency radius.}
    \label{fig:en_radius}
\end{figure}

Figure \ref{fig:coasting} presents the results of the coasting outflows, which are only driven for $5\times 10^4$~yr, after which the AGN shuts off forever. The panels and colours are identical to Figure \ref{fig:wabs_1}. 

The coasting outflow keeps expanding for more than an order of magnitude longer than the AGN episode duration. In the lowest density case (black solid line), the outflow stalls only at $t_{\rm stall} \simeq 4\times10^6$~yr, i.e. the outflow lifetime is a factor $80$ higher than that of the AGN. In the highest density case, the lifetime is $t_{\rm stall} \simeq 2\times 10^6$~yr, i.e. $40$ times longer than the AGN. It is important to note that even though the outflow keeps expanding, it does so very slowly: the stalling radius is $R_{\rm stall} = 0.3--0.9$~kpc in all cases, and the outflow spends $>80\%$ of its lifetime at $R > R_{\rm stall}/2$. 

The optical depth of the outflowing material depends on gas fraction. The radii at which transparency is reached are the same as in the continuous outflow models, i.e. $10 {\rm pc} < R_{\rm tr} < 200 {\rm pc}$. This range falls in the middle of the range of radii of observed WAs, which have $0.01 {\rm pc} < R_{\rm WA} < 10^7 {\rm pc}$, with the majority estimated to lie at $1 {\rm pc} < R_{\rm WA} < 100 {\rm pc}$ \citep{Tombesi2013MNRAS}. In the lowest-density case, the outflow becomes transparent very early, at $t < t_{\rm q}$, and drops to an optical depth as low as $\tau_{\rm min} = 10^{-2}$; the outflow in the densest case becomes only marginally optically thin ($\tau_{\rm min} \simeq 0.65$) during its evolution. In terms of column densities, the least dense model goes down to $N_{\rm gas} \sim 1.5 \times 10^{22}$~cm$^{-2}$, while the densest one stays at or above $N_{\rm gas} \sim 10^{24}$~cm$^{-2}$. These densities are $1-2$ orders of magnitude higher than observed: $10^{20} {\rm cm}^{-2} < N_{\rm H,obs} < 10^{23} {\rm cm}^{-2}$. We propose an explanation for this discrepancy in Section \ref{sec:structure}.

In Figure \ref{fig:vel_radius}, we show the relationship between outflow velocity and radius for the same simulations. Initially, the gas velocity can be as high as $2\times 10^3$~km~s$^{-1}$. At the radius of transparency, the velocity has dropped to $300$~km~s$^{-1} < v < 1000$~km~s$^{-1}$. For the majority of their evolution, the outflows move with $v < 400$~km~s$^{-1}$. Therefore, an outflow is most likely to be observed moving with a velocity of a few hundred km/s - this is consistent with observed WA properties. The effective energy loading factor, defined as $f_{\rm E} = \dot{E}_{\rm out} / L_{\rm Edd}$, is initially much higher than in energy-driven outflows \citep[where $f_{\rm E} \simeq 0.02$, cf.][]{Zubovas2018MNRAS}, due to much more efficient driving. At the radius of transparency, the energy loading factor is $0.05 < f_{\rm E} < 0.1$, some 2-3 orders of magnitude higher than that in observed WAs \citep{Tombesi2013MNRAS,Laha2016MNRAS}. It should be noted that our values of $f_{\rm E}$ are strict upper limits, since they include the kinetic energy of all the outflowing material, and the system is assumed to be perfectly adiabatic. Neither assumption is correct in reality; in particular, a very large fraction of the outflowing mass may be confined to cold clouds that are not observed as WAs (see Section \ref{sec:structure}). Very crudely, we estimate the observationally-derived energy loading factor to be a factor $N_{\rm H}/\left(1.5\times10^{24}{\rm cm}^{-2}\right)$ times the model-derived one, since this factor represents the ratio of observed WA mass to total outflowing mass. The observed WA column densities give a ratio of $10^{-4} - 0.1$, bringing the energy loading factors in line with observed ones. The radial trend of $f_{\rm E}$ decreasing with radius is also consistent with observations \citep{Tombesi2013MNRAS}.

The relation between velocity and energy loading factor in our models is very simple: $f_{\rm E} \propto \dot{E}_{\rm out} \propto \dot{M}_{\rm out} v_{\rm out}^2 \propto v_{\rm out}^3$. This relation is much steeper than the observed correlation \citep{Tombesi2013MNRAS}. Two effects may decrease the slope and bring it closer to observations, however. First of all, denser outflows have higher energy loading factors at the same velocity, but they can only be observed at lower velocities, once they reach $R_{\rm tr}$. Connecting the $f_{\rm E}$ and $v_{\rm out}$ values at the transparency radius for the different outflows leads to the relation having a shallower slope $f_{\rm E,tr} \propto v_{\rm out,tr}^{2.4}$. Secondly, the photon leakage mentioned above is more important when the energy loading factor is higher, because once $f_{\rm E} \simlt 0.02$, the outflow transitions to being wind-driven. Leaking photons result in lower effective coupling between AGN luminosity and outflow kinetic energy, but the mass flow rate, which depends on the solid angle the outflowing material covers, is reduced more strongly than velocity, which is effectively independent in every direction. This effect would bring the slope of the $f_{\rm E} - v_{\rm out}$ relation closer to $2$ for each gas fraction, and potentially to less than $2$ when the effect of transparency is considered.

The long-term evolution of outflows depends significantly on the velocity dispersion of the bulge as well. For example, if the bulge of the galaxy was twice as large, i.e. $R_{\rm b} = 2$~kpc, all coasting outflow would have a stalling time $t_{\rm stall} > 10^6$~yr and all galactic nuclei would become transparent very rapidly. Therefore, the condition for maintaining obscuration of the SMBH is that the galaxy is both compact and very gas-rich, as may be intuitively expected.

\section{Discussion} \label{sec:discuss}

\subsection{Radii of warm absorbers}

We saw above that it takes about 40-80 times the duration of the AGN event for the outflow to reach the stalling radius $R_{\rm stall}$. Since the duty cycle of AGN is of order a few percent \citep{Schawinski2010ApJ}, this is roughly the same as the expected average time between successive AGN events. So it is likely that before the outflow can collapse, it is hit by AGN disc wind from a subsequent episode. The stalled (or almost stalled) outflow shell provides a natural barrier to the wind, producing a shock at a radius $R_{\rm sh} \lesssim R_{\rm stall}$. Over time, the position of the barrier can shift, depending on the activity history of the galaxy. If the galaxy has a higher-than-average duty cycle, the material might be pushed outward, while in the opposite case, the material can fall back to lower radii.

The transparency radius is positively correlated with the galaxy gas fraction, while the stalling radius is anti-correlated (see Figure \ref{fig:coasting}, top panel). This difference means that we unfortunately cannot predict a correlation between warm absorber radius and galaxy gas fraction, since we don't know whether warm absorbers form close to the transparency radius, or close to the stalling radius. On the other hand, if a subset of warm absorbers with a narrow range of column densities is analysed, their radii should be positively correlated with galaxy gas mass and/or gas fraction.

\subsection{Thermal and ionization structure} \label{sec:structure}

The stalled or slowly moving shell presents a barrier to any subsequent winds blowing from the AGN. As the wind shocks against the shell, it heats up to temperatures $T_{\rm sh} > 10^{10}$~K, but then cools down rapidly via inverse Compton and free-free processes \citep{King2015ARA&A}. The shocked wind region extends from very close to the AGN $\left(R_{\rm sh} \sim 10^{17} {\rm cm}\right)$ out to the shell radius. The cooling timescale is very sensitive to distance from the AGN: $t_{\rm cool} \propto R^2$. So the shocked wind probably coalesces into clumps of various sizes from the inside out. The coldest clumps are the densest, and we expect an anti-correlation between distance from AGN and ionization parameter. Clump temperature, assuming ionization equilibrium, is directly proportional to ionization parameter for a wide range of the latter \citep{Sazonov2005MNRAS}, so the presence of different ionization species should also depend on distance to the AGN. Together, these clumps produce the observed WA spectra in the distance range consistent with observations \citep{Tombesi2013MNRAS, Laha2016MNRAS}. Shocked wind has been proposed as an explanation for several individual WAs \citep{Blustin2005A&A, Pounds2013MNRAS, Sanfrutos2018arXiv}.

The shocked wind also drives a forward shock into the slow-moving shell. Due to the vast difference of densities, the forward shock is rather slow, with velocities of order a few to several tens of kilometres per second. It is plausible that a significant fraction of the shell material is heated to a temperature $T \sim 10^4$~K, from which it rapidly cools down and reaches the ``opacity gap'' \citep[e.g.,][]{Bell1994ApJ}. This material becomes transparent to soft photons, reducing the opacity of observed WAs. Opacity to hard photons is reduced by another outcome of cooling - gas fragmentation (see also Sections \ref{sec:feeding} and \ref{sec:sf} below). The dense gas shell forms very dense clumps, while the column density of the inter-clump material becomes much lower than the average shell column density. This may be the reason for the discrepancy in column densities between our models and observation results (see Section \ref{sec:coasting}. We predict that future observations will show WAs accompanied by almost co-spatial atomic gas reservoirs and very dense clouds with masses exceeding those of the absorber by an order of magnitude or more. The existence and position of these clouds may be determined with observations of light echoes, but this may require long-term monitoring of individual AGN.

\subsection{Compton-thick AGN}

A significant part of the total accretion energy released during the build-up of the SMBH population may be heavily obscured \citep{Comastri2004ASSL,Lansbury2017ApJ,Tasnim2018arXiv}. Compton-thick AGN have hydrogen column densities $N_{\rm H} > 1.5\times10^{24}$~cm$^{-2}$ and, by definition, a Compton scattering optical depth $>1$. 

In our picture, most AGN experience such a phase before their outflows clear out the immediate environment, thus reducing the obscuring column. The length of this phase and the number of such phases experienced over the lifetime of each galaxy depend on galaxy velocity dispersion, gas fraction and AGN accretion history. Specifically, in compact gas-rich galaxies, the Compton-thick phase would last longer, and may persist for numerous AGN episodes, since more than one episode's worth of energy is required in order to push gas out to the transparency radius. Depending on the accretion history of an individual AGN, it may move back into a Compton-thick state some time after reaching transparency, as the outflow stalls and falls back to the centre. Furthermore, if the outflowing material becomes self-gravitating (see Section \ref{sec:sf} below), the distribution of optical depths along different lines of sight becomes more uneven, leading to a variety of observed column densities in otherwise similar galaxies.

We therefore predict that Compton-thick AGN should be more common in the early Universe, and overall in more gas-rich and compact galaxies with smaller central black holes. Those black holes would then on average accrete more rapidly than the general population (see Section \ref{sec:feeding} below). This scenario is consistent with the finding that Compton-thick AGN probably host the smallest and most rapidly growing SMBHs \citep{Goulding2011MNRAS, Lanzuisi2015A&A}, and that Compton-thick AGN are found in the most gas-rich systems \citep{Nardini2011MNRAS}. They are also frequently found in merging systems \citep{Kocevski2015ApJ,Lansbury2017ApJ}, which are likely to have more gas in the central regions.

Light echoes may be used to distinguish whether a system with observed WAs has recently been Compton-thick. The narrow line region of an AGN that was very recently Compton-thick should show evidence of being ionized by very hard photons, while the centre would show ionization by both hard and soft photons. If Compton-thick AGN and WAs form an evolutionary sequence, then we would expect systems with Compton-thick light echoes to be predominantly those that have WAs, while if these two populations are unrelated, no correlation should appear.

\subsection{Implications for AGN feeding} \label{sec:feeding}

The transparency and stalling radii are much larger than the size of accretion discs ($R_{\rm d} \ll 1$~pc) and larger than the typical sizes of molecular clouds ($R_{\rm cl} \sim 1-10$~pc) that might feed the SMBH. The optically thick outflow should destroy whatever large-scale reservoir is feeding the SMBH. By the time the AGN becomes visible to distant observers, the central engine is only powered by what remains of its accretion disc, truncated at the self-gravity radius $R_{\rm sg} \sim 0.01$~pc \citep{King2007MNRAS}. This result ties in with two important aspects of AGN evolution. 

First, the typical observationally-inferred AGN episode duration, $t_{\rm q} \sim 10^5$~yr \citep{Schawinski2015MNRAS}, is very similar to the analytically-derived timescale of accretion disc consumption \citep{King2015MNRAS}. This is only possible if, once an AGN episode begins, the accretion disc is consumed without being refilled. Destruction of any large-scale reservoir that feeds the SMBH or might feed it in the near future ensures this, and explains why we apparently do not observe extremely long AGN episodes.

Second, the existence of tight correlations between AGN luminosity and large-scale outflow properties implies that multiple AGN episodes in a single galaxy tend to have similar maximum luminosities, i.e. similar Eddington ratios $L_{\rm AGN}/L_{\rm Edd}$ \citep{Zubovas2018MNRAS}. Since any large-scale reservoir is destroyed by the expanding radiation-pressure-driven bubble, the maximum luminosity of the AGN episode cannot depend strongly on the properties of the gas reservoir that fed the accretion disc, but only on the properties of this disc itself, which is limited by self-gravity. So each AGN episode in an individual galaxy should have the same maximum luminosity. In some cases the luminosity of previous AGN episodes can be estimated from light echoes or dynamical footprints, and it may be possible to determine the properties of two (or even more) consecutive AGN episodes. This would allow an observational test of this prediction.

Another important implication is that the shocked gas shell is unstable in two ways. Since it is pushed from the centre by a much less dense shocked wind, it is RT-unstable \citep{King2010MNRASb} and will form dense fingers falling back toward the SMBH. Secondly, since it is formed of material compressed from a sphere of radius $R$ to a shell of thickness $\Delta R < R$, it can become gravitationally unstable as long as $\Delta R / R < f_{\rm g}$. Especially in gas-rich galaxies, this may lead to powerful bursts of star formation (see Section \ref{sec:sf} below). Self-gravity also increases the density contrasts between different regions of the shell, exacerbating the effects of RT instability. The dense gas clumps detaching from the shell and falling on to the SMBH can feed it very efficiently, at rates close to the dynamical rate
\begin{equation}
\dot{M}_{\rm dyn} = \frac{f_{\rm g} \sigma^3}{G},
\end{equation}
which can exceed the Eddington rate for a black hole of mass
\begin{equation}
M_{\rm BH} < \frac{f_{\rm g} \kappa \eta c \sigma^3}{4 \pi G^2} = \frac{\eta c f_{\rm g}}{4 \sigma f_{\rm c}} M_\sigma.
\end{equation}
The final expression is $> M_\sigma$ as long as $f_{\rm g}/f_{\rm c} \simgt 0.03 \sigma_{200}$, which is definitely the case for these buried AGN. Therefore SMBHs of any mass can be fed at super-Eddington rates via this mechanism. This scenario is similar to the ``black hole foraging'' \citep{Dehnen2013ApJ}, where feedback episodes enhance orbit-crossing of different gas streams around a SMBH, leading to enhanced accretion rates. Here, the main difference is that the system is gas-rich, and orbit-crossing is not necessarily required. The effect of AGN feedback is to ``churn'' the surrounding material, significantly enhancing the importance of radial motions due to repeated blow-out and re-accretion of material. 

The (mildly) super-Eddington growth of black holes in gas-rich systems may also explain the origin of the first SMBHs, observed to have masses $M_{\rm BH} > 10^9 \; \msun$ at $z > 6$ \citep[e.g.][]{Mortlock2011Natur, Venemans2013ApJ}. In order for these black holes to grow to such masses in such a short time after the Big Bang, they have to either start from very massive seeds \citep{Begelman2006MNRAS}, have very low spins which lead to very low radiative efficiencies \citep{King2006MNRAS}, or grow at super-Eddington rates for at least a fraction of time \citep[and references therein]{Madau2014ApJ,Pezzulli2016MNRAS}; see also \citet{Latif2016PASA} for an overview of the proposed models. A dense gas shell, continuously churned by AGN feedback episodes, provides a reservoir of gas that can feed the SMBH for prolonged periods of time, facilitating its growth at early cosmic times. We note that a similar scenario of heavily-obscured black holes in the early Universe was recently proposed as an explanation for strength of the 21 cm absorption signal from the first stars \citep{Ewall2018ApJ}; our model is consistent with this hypothesis.

\subsection{Star formation in galactic cores} \label{sec:sf}

The self-gravitating gas clumps forming in the compressed shell (see above) result in star formation. Although the details depend on many factors, most importantly the gas fraction and heating by repeated wind shocks impacting the shell from within, the process may be significant. The total mass of the shell is $M_{\rm sh} \simeq f_{\rm g}^2/f_{\rm c} M_\sigma$, i.e. similar to or larger than the mass of the central black hole or even the mass it would eventually reach. The shocked ISM gas cools rapidly \citep{Zubovas2014MNRASb,Richings2018MNRAS,Richings2018MNRASb}, so a significant fraction of the shell can be converted into stars, at least until the remaining gas is too dilute to be self-gravitating. This leads to formation of a nuclear stellar cluster and/or a significant addition to the stellar population of the galactic bulge, depending on the stalling radius of the shell. Star formation and cooling do not explicitly depend on the radius of the shell \citep{Richings2018MNRASb}, therefore the relative importance of star formation versus SMBH feeding as sink terms for the gas should be roughly similar on different physical scales. However, cooling is more efficient at higher densities, therefore star formation should be faster overall in gas-rich galaxies with small central black holes, which produce outflows with small stalling radii. As a result, the initial formation of a nuclear star cluster should be affected significantly more by the AGN outflows than later growth of the bulge. This process may to some extent explain the observed faster growth of galactic bulges compared with SMBHs in the early Universe \citep{Greene2010ApJb}. At later times, when the shell around the SMBH is no longer self-gravitating, vigorous star formation ceases and the SMBH can catch up to the bulge. However, occasional episodes of star formation can occur simultaneously with super-Eddington accretion episodes \citep{Nayakshin2018MNRAS}.

\section{Conclusions} \label{sec:concl}

We used 1D numerical integration to investigate the evolution of spherically symmetric outflows driven by AGN buried in dense reservoirs of gas. The gas makes an optically thick envelope which scatters or absorbs all the AGN luminosity, leading to a more powerful outflow than in the standard wind-driven case, where only $\sim 5\%$ of the AGN luminosity drives the outflow. A typical AGN episode lasting for $5\times 10^4$~yr injects enough energy into the gas to drive the outflow bubble to approximately the transparency radius, defined by $\tau\left(R_{\rm tr}\right) = 1$, even in compact galaxy bulges with high velocity dispersion $\sigma \simeq 300$~km~s$^{-1}$; in less compact bulges, a single AGN episode may push gas much further out than $R_{\rm tr}$. The outflow slows down once the AGN episode ends, and moves with a modest radial velocity for a significant period of time $t_{\rm stall} \sim 1$~Myr. This produces a dense gas barrier for subsequent AGN winds to shock against. Cooling of these shocked winds results in spectral line and continuum radiation consistent with that of warm absorbers; the typical radius of $0.1-1$~kpc and velocity  of $10^2-10^3$~km~s$^{-1}$ of the shell is consistent with WA properties as well. In gas-rich compact galaxies, the shell is likely to be Compton--thick along many sightlines, suggesting an origin for this class of AGN. Significantly, the total mass in the dense shell is $M_{\rm out} \simeq f_{\rm g}^2/f_{\rm c} M_\sigma$, comparable with the final $M - \sigma$ mass the SMBH will eventually reach.

The existence of such a dense shell suggests that AGN can grow very rapidly in dense gas reservoirs at high redshift, by repeatedly agitating this surrounding gas shell and feeding efficiently from the unstable gas clumps that fall inwards from it. This process may be difficult to observe since the AGN may be Compton--thick. The shell can also fragment to form stars, building up a nuclear star cluster and a significant part of the galaxy bulge along with the growth of the SMBH.

\section*{Acknowledgements}

KZ is funded by the Research Council Lithuania grant no. MIP-17-78.
ARK acknowledges support from the UK STFC.





\bsp	
\label{lastpage}
\end{document}